\documentclass[twocolumn,showpacs]{revtex4}

\usepackage{graphicx}

\begin{document}

\newcommand{\beq}{\begin{equation}}
\newcommand{\eeq}{  \end{equation}}
\newcommand{\bea}{\begin{eqnarray}}
\newcommand{\eea}{  \end{eqnarray}}
\newcommand{\tr}{{\rm tr}}

\title{Random matrix ensembles from nonextensive entropy}

\author{Fabricio Toscano} 
\email{toscano@if.ufrj.br}
\affiliation{Universidade Federal do Rio de Janeiro (UFRJ), 
             Instituto de F\'{\i}sica, Caixa Postal 68528, \\ 
             21941-972, Rio de Janeiro, Brazil}
\author{Ra\'ul O. Vallejos} 
\email{vallejos@cbpf.br}
\homepage{http://www.cbpf.br/~vallejos}
\author{Constantino Tsallis} 
\email{tsallis@cbpf.br}
\homepage{http://tsallis.cat.cbpf.br/biblio.htm}
\affiliation{Centro Brasileiro de Pesquisas F\'{\i}sicas (CBPF),
	       Rua Dr.~Xavier Sigaud 150, \\
             22290-180 Rio de Janeiro, RJ, Brazil}

\date{\today}

\begin{abstract}
The classical Gaussian ensembles of random matrices can be constructed 
by maximizing Boltzmann-Gibbs-Shannon's entropy,
$ S_{\rm BGS} = - \int d{\bf H} [P({\bf H})] \ln [P({\bf H})],$
with suitable constraints. 
Here we construct and analyze random-matrix ensembles arising from the 
generalized entropy 
$S_q = \left( 1 - \int d{\bf H} [P({\bf H})]^q \right)/(q-1)$
(thus $S_1=S_{\rm BGS}$).
The resulting ensembles are characterized by a parameter $q$ measuring
the degree of nonextensivity of the entropic form. 
Making $q \to 1$ recovers the Gaussian ensembles.
If $q \ne 1$, the joint probability distributions $P(\bf H)$ 
cannot be factorized, i.e., the matrix elements of $\bf H$ are correlated.
In the limit of large matrices two different regimes are observed. 
When $q<1$, $P(\bf H)$ has compact support,
and the fluctuations tend asymptotically to those of the Gaussian ensembles.
Anomalies appear for $q>1$: Both $P(\bf H)$ and the marginal distributions 
$P(H_{ij})$ show power-law tails. Numerical analyses reveal that 
the nearest-neighbor spacing distribution is also long-tailed (not Wigner-Dyson) 
and, after proper scaling, very close to the result for the $2 \! \times \! 2$ 
case -- a generalization of Wigner's surmise. We discuss connections
of these ``nonextensive" ensembles with other non-Gaussian ones, like
the so-called L\'evy ensembles and those arising from soft-confinement.
\end{abstract}





\maketitle

\section{Introduction}
\label{sec1}

The Gaussian ensembles of Random Matrix Theory provide
the standard statistical description of spectral fluctuations 
in a multiplicity of quantum systems ranging from nuclei to disordered 
mesoscopic conductors and classically chaotic systems 
\cite{porter65,mehta67,brody81,bohigas91,guhr98}. 

Gaussian ensembles can be obtained from two postulates: 
the invariance of the joint distribution probability $P(\bf H)$ 
with respect to changes of bases and the statistical independence 
of matrix elements. An alternative and more appealing way of constructing
random matrix ensembles uses a maximum entropy principle 
\cite{porter65,balian68}. 
One constraint is normalization,
\beq
\label{norm}
\int d {\bf H} \, P({\bf H}) = 1 \;.
\eeq
The other one has the purpose of confining the spectrum, 
but is otherwise arbitrary (as long as the integral converges),
\beq
\int d {\bf H} \, P({\bf H}) \, \tr  V({\bf H}) = 1 \;
\eeq
(the trace ensures rotational invariance). 
For instance, the Gaussian ensemble of real symmetric
matrices is obtained by the simplest choice
\beq
\label{quad}
V({\bf H}) = {\bf H}^2 \;.
\eeq
It has been proven that, in the limit of large matrices, 
and for a strong enough confining potential $V$, local
fluctuation properties tend to those of the Gaussian case,
whatever the shape of $V$ \cite{brezin93,hackenbroich95}.

To escape from Gaussian universality one must consider
soft-confinement potentials \cite{muttalib93,bogomolny97},
or breaking rotational invariance. The latter case typically 
arises when matrix elements $H_{ij}$ are independent (non Gaussian) 
random variables.
For instance, Cizeau and Bouchaud constructed anomalous 
``L\'evy ensembles" by drawing $H_{ij}$ from a long-tailed 
distribution \cite{cizeau94}.

The purpose of this paper is to present a new way of constructing
non-Gaussian ensembles while preserving rotational invariance. 
The idea is to use a maximum entropy approach with the 
usual constraints but with the nonextensive entropy \cite{tsallis}:
\beq
\label{tsallis}
S_q [P({\bf H})] = \frac{ 1-\int d{\bf H} \; [P({\bf H})]^q }
                        { q-1 } \; ,
\eeq
where $q$ is a free real parameter ($q=1$ recovers Shannon's 
standard entropy). This scheme produces a variety of 
ensembles, with $q$ controlling the degree of confinement. 
Some ensembles belong to the Gaussian universality class but 
others exhibit anomalous behavior, characterized by distributions 
having power-law tails.

The explicit construction of these $q$-ensembles is presented
in Sect.~\ref{sec2}, where we also derive expressions for marginal
distributions and the joint density of eigenvalues. 
Remarkably, for large matrices, the $q$-ensembles can be 
represented as a superposition of Gaussian ensembles. 
This allows us to obtain closed analytical formulas for the eigenvalue
density, level-spacing probability distributions, etc (Sect.~\ref{sec3}). 
The comparison of analytical results with numerical simulations is
the subject Sect.~\ref{sec4}. We present in Sect.~\ref{sec5} the 
concluding remarks.

\section{The generalized ensembles}
\label{sec2}

For simplicity we restrict our analysis to ensembles of real and 
symmetric matrices $\bf H$ -- extensions are straightforward.
The volume element in this space is 
\beq
d\mbox{\bf H} =
\prod_{i=1}^N \; dH_{ii} \; 
\prod_{i<j}^N \; dH_{ij} \; ,
\eeq
where it is understood that matrices are of size $N \times N$.
Generalized ensembles are obtained by maximizing the entropy 
of Eq.~(\ref{tsallis}) subjected to normalization, Eq.~(\ref{norm}), 
and
\beq
\label{second}
\frac{\int d\mbox{\bf H} \; \tr {\bf H}^2 \; [P(\mbox{\bf H})]^q} 
{\int d\mbox{\bf H} \; [P(\mbox{\bf H})]^q} = \sigma^2 \;,
\eeq
with $\sigma$ a constant having units of energy (we are assuming that
of $\bf H$ is a Hamiltonian).
Equation~(\ref{second}) is the generalization of the usual constraint
that leads to the Gaussian ensembles in the standard maximum entropy
approach. Arguments justifying the use of the escort probabilities $P^q$,
and applications of this generalized maximum entropy scheme to various
problems can be found in Ref.~\cite{tsallis}. 

Using the Lagrange multiplier technique it is straightforward to 
find the distribution of maximum entropy:
\beq
\label{joint}
P({\bf H}) \propto 
\exp_q \left( -\lambda \; \tr {\bf H}^2 \right) \;,
\eeq
where we have defined the $q$-exponential function \cite{tsallis}
\beq
\label{expq}
\exp_q (x) \equiv \left\{ [1+(1-q)x]_+ \right\}^{1/(1-q)} \; ,
\eeq
with
\beq
\label{bracket}
\&_+ = \max\{\&,0\} \; 
\eeq
[note that $\exp_1(x)=\exp(x)$].
The omitted normalization constant in (\ref{joint}) and the parameter $\lambda$ 
can be determined from the constraints (\ref{norm}) and (\ref{second}).
(Some preliminary results along these lines have been obtained by 
Evans and Michael \cite{evans02}).

The ensemble defined by Eq.~(\ref{joint}) will be called 
the ``$q$-Orthogonal Ensemble" (qOE), as it can be seen in (\ref{joint}) 
that the probability distribution depends only on $\tr {\bf H}^2$, 
an orthogonal invariant. 
When $q \to 1$ the $q$-exponential function tends to the usual 
exponential,
and one recovers the Gaussian Orthogonal Ensemble (GOE).
Except for the $q=1$ case, the $q$-exponential in (\ref{joint}) 
{\em cannot} be factorized into a product of 
(marginal) distributions for individual matrix elements 
$H_{ij}$, which are then correlated.
We can already verify that the cases $q<1$ and $q>1$ are qualitatively
different. Equations (\ref{expq}) and (\ref{bracket}) show that for $q<1$ 
the distributions have compact support; if $q>1$, there are always power-law
tails (we are assuming $\lambda>0$, see below). 

To proceed with the analysis of qOE it will be convenient to think of 
matrices $\bf H$ as points in a $d$-dimensional euclidean space \cite{porter65,lecaer99,delannay00}. 
The first $N$ components of a point $\bf r$ correspond to diagonal elements 
$H_{ii}$, the last ones to the upper triangle $H_{ij}$, $i<j$:
\beq
\label{map}
{\bf r} = 
(H_{11},\cdots,H_{NN},\sqrt{2} H_{12},\cdots,\sqrt{2} H_{N-1,N}) 
\; .
\eeq
The dimension of this space equals
the number of independent matrix elements of $\bf H$, i.e.,
\beq
d = \frac{N(N+1)}{2} \; .
\eeq
The scaling of $H_{ij}$ by $\sqrt{2}$ makes the probability 
distribution $(\ref{joint})$ spherically symmetric in $R^d$, 
i.e., $P_{\rm qOE}({\bf r})$ is the product of a uniform distribution 
in the angles, and a radial distribution \cite{omission}
\beq
\label{pofr}
{\cal P}(r;q,\sigma,N) 
\propto r^{d-1} \exp_q \left( -\lambda r^2 \right) \; ,
\eeq
where
\beq
r^2 \equiv {\bf r} \cdot {\bf r} = \tr {\bf H}^2 \; .
\eeq
The observations above imply that qOE belong to the wider category 
of ``spherical ensembles" recently studied by Le Ca\"er and Delannay
\cite{lecaer99,delannay00}. 

For $q>1$, the distribution (\ref{pofr}) has a power-law tail that 
goes like $1/r^{(1+\mu)}$ with
\beq
\label{mu}
\mu = \frac{2}{q-1} - d   \; .
\eeq
Then the normalization condition cannot be satisfied for all values 
of $q$, but only by those making $\mu>0$, i.e., 
\beq
\label{subset}
-\infty < q < 1+ \frac{2}{d}\;\;.
\eeq
(Note the formal similarity between this problem and the 
generalized random walker in $d$ dimensions \cite{anomalous}.)

The Lagrange multiplier $\lambda$ is given by
\beq
\label{lambda}
\lambda = \frac{1}{\sigma^2} \; \frac{d}{2- d \,(q-1)} \; .
\eeq  
Inside the region (\ref{subset}) (i.e., normalizability) 
$\lambda$ is always positive.

Integrating Eq.~(\ref{joint}) over all variables but one, 
we obtain the marginal distributions for diagonal and 
off-diagonal matrix elements, \cite{mendes01}  
\bea
\label{marginal1}
P(H_{ii}) & \propto & 
\exp_{q'} \left( -   \lambda' H_{ii}^2 \right)  \; , \\
\label{marginal2}
P(H_{ij}) & \propto & 
\exp_{q'} \left( - 2 \lambda' H_{ij}^2 \right)  \; ,
\eea
where
\beq
q' = \frac{2-(d-3)(q-1)}{2-(d-1)(q-1)} \; 
\eeq
and
\beq
\lambda' = 
\frac{d}{2\sigma^2} \; \frac{2-(d-1)(q-1)}{2-d\,(q-1)} \; .
\eeq
The following properties can be easily verified.
The parameter $q'$ is an increasing function of $q$, and
around the critical value $q=1$ one has 
\beq
q' = q + {\cal O}\left[(q-1)^2\right] \; .
\eeq
In addition, $\lambda'$ is always positive.
Then, in parallel with the global $P({\bf H})$, the marginal 
distributions also decay as power laws or have compact 
support, depending on $q$ being larger or smaller than 
one, respectively. We remark that the matrix elements are
not independent, so, $P({\bf H})$ cannot be reconstructed 
from the marginal probabilities 
(\ref{marginal1},\ref{marginal2}).

The joint density of eigenvalues can be obtained in a 
straightforward way: \cite{bohigas91,evans02}
\beq
P(\varepsilon_1,\cdots,\varepsilon_N) \propto
\prod_{i<j=1}^N \left| \varepsilon_j-\varepsilon_i \right|
\exp_q \left( -\lambda \sum_{i=1}^N \varepsilon_i^2 \right) \; .
\eeq
The part that is responsible for level repulsion is identical
to that in GOE because it arises only from orthogonal symmetry.
The difference is in the confinement term, which in the present case 
is a non-separable $q$-exponential.
Thus the ``potential" that confines the spectrum is not 
a single-particle quadratic well, as in GOE. It is rather a 
mean field, proportional to the moment of inertia
$\sum \varepsilon_i^2$.

We can get a clear view of the generalized ensembles
by noting that these are connected to the so-called Fixed Trace
Ensembles (FTE) and, for large $N$, to the Gaussian ensembles. 
In fact, recall that FTE are defined by \cite{porter65,lecaer99,delannay00}
\beq
P_{\rm FTE} ({\bf H}; r, N) 
\propto \delta (\tr {\bf H}^2 - r^2) \; .
\eeq
Let $f({\bf H})$ be an arbitrary function and consider
the averages in both ensembles qOE and FTE, namely,
\beq
\langle f({\bf H}) \rangle_{\rm qOE} (q,\sigma,N) = 
\int d{\bf H} P_{\rm qOE}({\bf H};q,\sigma,N) f({\bf H})   \; , 
\eeq
and 
\beq
\langle f({\bf H}) \rangle_{\rm FTE} (r,N)        
\int d{\bf H} P_{\rm FTE}({\bf H};r,N) f({\bf H})   \; .
\eeq
Then we have the relation
\bea
& & 
\langle f({\bf H}) \rangle_{\rm qOE} (q,\sigma,N) = \nonumber \\
& &
\int_0^\infty dr \; {\cal P}(r;q,\sigma,N) 
\langle f({\bf H}) \rangle_{\rm FTE} (r,N) \; .
\label{relation}
\eea
The average over qOE can be calculated in two stages. 
First do the average over the angles, for a fixed radius $r$. 
This correspond to a FTE average. 
Then average over radii, with the weighting function ${\cal P}(r)$.
Of course, the same is true for the GOE, which corresponds to
the particular case $q=1$. The relationship between 
qOE (or GOE) and FTE is analogous to that between the canonical 
and microcanonical ensembles of Statistical Mechanics. 

Equation (\ref{relation}) involves no approximations. 
Although exact, it is not very useful because it requires the knowledge 
of fixed-trace averages. However, if one is interested in the limit
of large matrices, important simplifications can be 
made.

\section{The large $N$ limit}
\label{sec3}

The key point is that, for $N$ large enough, the FTE average 
in the r.h.s. of (\ref{relation}) can be approximated by an average
in a GOE having the property $\langle \tr {\bf H}^2 \rangle =r^2$.
Then, if we know the GOE average of a given function,
its corresponding qOE average can in principle be calculated
by doing just one integration. We will analyze in detail two
spectral statistics: the eigenvalue density,
\beq
\label{rho}
\rho (\varepsilon; q , \sigma, N) = 
\left \langle
\sum_{i=1}^N \delta (\varepsilon - \varepsilon_i) 
\right \rangle \; ,
\eeq
and the distribution of level spacings,
\beq
\label{pofs}
p(s ; q , \sigma, N) = 
\left \langle 
\delta( \varepsilon_{i+1} - \varepsilon_{i} - s ) 
\right \rangle \; .
\eeq
In the last equation $\varepsilon_i$ and $\varepsilon_{i+1}$ 
are two consecutive eigenvalues lying at the center of the band, 
i.e., $\varepsilon_i \approx 0$. 
It is (or will become) clear that other statistics, e.g., two-point correlation 
functions, can also be considered along the same lines.

In order to obtain the qOE averages of (\ref{rho}) and (\ref{pofs})
we need the corresponding FTE expressions, to be further averaged 
with ${\cal P}(r;q,\sigma,N)$, as indicated by Eq.~(\ref{relation}). 
However, we will approximate FTE averages by the corresponding GOE ones. 
Then, the basic ingredients become the ``semicircle law" (for the
eigenvalue density),
\beq
\label{semicircle}
\rho (\varepsilon;N,r) =
\frac{N^2}{2\pi r^2} 
\sqrt{ \frac{4r^2}{N} - \varepsilon^2} \; ,
\eeq
and Wigner's surmise, 
\beq
\label{Wigner}
p(s;N,r) = 
\frac{N^3 s}{2\pi r^2} \exp \left( -\frac{N^3 s^2}{4 \pi r^2} \right) \; ,
\eeq
giving the level-spacing distribution. Equations~(\ref{semicircle}) 
and (\ref{Wigner}) are good approximations
for both GOE and FTE distributions when $N$ is large \cite{lecaer99}. 

We recall that if $q>1$, the normalization condition (\ref{subset})
limits the value of $N$ to a finite domain. 
On the other hand, the case $q<1$ does not present such a problem.
So, we analyze both cases separately.

\subsection{ Ensembles with $q<1$ }
\label{sec3a}

Except for providing an energy scale, $\sigma$ plays no special 
role. From now on, without loss of generality, we set $\sigma=1$.
If desired, $\sigma$ can be restored at any time by dimensional 
analysis.

When $N \to \infty$ ($q$ fixed) the radial distribution of qOE tends to 
\beq
{\cal P}(r;q,N) \propto r^{d-1} \left[ 1-r^2 \right]^{1/(1-q)} \; ,
\eeq
limited to the domain $0 \le r \le 1$. As $d$ grows, the distribution 
is squeezed against $r=1$, being concentrated in a small 
region below $r=1$, of width ${\cal O}(1/d)={\cal O}(1/N^2)$. 
It can be verified that both the level density (\ref{semicircle}) 
and the spacing distribution (\ref{Wigner}), when considered as functions 
of $r$, have widths which are ${\cal O}(1)$. 
Thus the radial distribution is much narrower and we can safely 
approximate 
\beq
\label{delta}
{\cal P}(r; q<1, N\to \infty) \simeq \delta(r-1) \; .
\eeq
We conclude that, when $q<1$ and $N \to \infty$ the ensembles qOE 
tend to the GOE [as far as it concerns the distributions being
studied, namely Eqs.~(\ref{rho}) and (\ref{pofs})].

\subsection{ Ensembles with $q>1$ }
\label{sec3b}

\begin{figure}[!hbt]
\includegraphics[angle=-90,width=8cm]{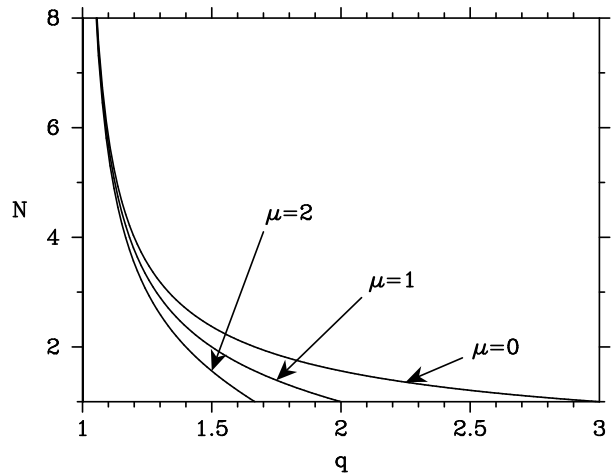} 
\caption{When $q>1$ the ensembles qOE lie in the region 
limited by the axes and the curve $\mu=0$ (normalization
frontier).
As $N$ becomes large, the maximum $q$ allowed tends to one. 
Lines correspond to families of ensembles having the same power-law 
tails (labeled by $\mu$).}
\label{fig1}
\end{figure}

When $q>1$ the possible ensembles are restricted to a 
region in the plane $q-N$ that gets thinner as 
$N \to \infty$ (see Fig.~\ref{fig1}). 
%
The natural coordinates in this region are $N$ (or $d$) and $\mu$ 
[see Eq.~(\ref{mu})], the latter controlling the tails of ${\cal P}(r)$ 
and other distributions. For instance, substituting (\ref{mu}) and (\ref{lambda}) 
into (\ref{marginal1}) or (\ref{marginal2}), one immediately verifies that the 
marginal distributions behave asymptotically like 
\beq
\label{levy}
P(H_{ij}) \sim \frac{1}{H_{ij}^{1+\mu}} \; .
\eeq
The radial distribution (\ref{pofr}), as a function of $r,\mu,N\/$ 
becomes:
\beq
{\cal P}(r,\mu,N) \propto r^{d-1} 
\left[ 1 + \frac{d}{\mu} \, r^2 \right]^{-(d+\mu)/2} \; .
\eeq
This expression allows the identification of some well 
known ensembles as special members
of the qOE class: the Cauchy-Lorentz ensemble corresponds to 
$\mu=1$. 
An integer $\mu >1$
produces Student's ensembles (see \cite{lecaer99,delannay00} 
and references therein; see also \cite{souza97}).

Now we analyze the limit $N \to \infty$ while keeping $\mu>0$ 
fixed, i.e., we move upwards along the curves of Fig.~\ref{fig1}. 
As in the case $q<1$, examined before, there is a limiting 
distribution. Some simple algebra leads to
\beq
\label{pofrmu}
{\cal P}(r,\mu, N \to \infty) \propto 
r^{-(1+\mu)} \exp \left( - \frac{\mu}{2r^2} \right)  \; .
\eeq
Only when $\mu \to \infty$, ${\cal P}$ tends to 
the delta function (\ref{delta}), and GOE is recovered.
For finite $\mu$ the width of ${\cal P}(r)$ is at least 
${\cal O}(1)$. In any case, the average of a given GOE
distribution with ${\cal P}(r)$ gives the corresponding
qOE distribution (via Eq.(\ref{relation}) with 
$\langle f({\bf H}) \rangle_{FTE} \sim \langle f({\bf H}) \rangle_{GOE}$, when $N \to \infty$). Let us first consider the density of
states. Inserting Eqs.~(\ref{semicircle}) and (\ref{pofrmu}) 
into (\ref{relation}) we obtain 
\beq 
\label{rhomu}
\rho (\varepsilon;\mu) \propto 
\int^\infty_{\sqrt{N} \varepsilon/2} dr  \,
\frac{ \sqrt{4r^2-N\varepsilon^2}}{r^{\mu + 3}}
\exp \left( - \frac{\mu}{2r^2} \right) \; .
\eeq
This integral can not be expressed in terms of elementary
functions. However, some information can be extracted 
analytically. Setting $\varepsilon=0$ one obtains the
qOE density of states at the center of the band, 
\beq
\label{rhozeromu}
\rho (0;\mu) = 
\frac{N^{3/2}}{\pi} 
\frac{ \Gamma \left[ (\mu+1)/2  \right] }
     { \Gamma \left[  \mu   /2  \right] } \sqrt{\frac{2}{\mu}} \; .
\eeq
The behavior for large $\varepsilon$ can be easily 
recognized by making the change of variables 
$2r=\sqrt{N} \varepsilon z$ in (\ref{rhomu}), which leads to
\beq
\label{rhoemu}
\rho (\varepsilon;\mu) \propto \varepsilon^{-(1+\mu)}
\int_1^\infty dz \, 
\frac{\sqrt{z^2-1}}{z^{\mu + 3}}
\exp \left( - \frac{2 \mu}{N z^2 \varepsilon^2} \right) \; .
\eeq
Evidently the tails vanish like $\varepsilon^{-(1+\mu)}$.
This is also the behavior observed by Cizeau and Bouchaud 
in their ``L\'evy ensembles" of matrices having independent 
entries distributed according to the same law of Eq.~(\ref{levy}) 
\cite{cizeau94}.
We note, however, that the analogies can not be pushed further 
because our ensembles are rotationally invariant and L\'evy 
ensembles are not (the ensembles of Ref.~\cite{cizeau94} belong to 
the so-called $\alpha$-symmetric class \cite{lecaer99}).

The calculation of the spacing distribution proceeds as before.
We have to insert Eqs.~(\ref{Wigner}) and 
(\ref{pofrmu}) into (\ref{relation}). The result is
\beq
\label{psmu}
p (s;\mu) \propto s 
\int_0^\infty dr \, r^{-( \mu + 3)}
\exp \left[ - \frac{\mu}{2r^2} 
             \left( 1 + \frac{N^3 s^2}{2 \pi \mu} \right) 
    \right] \; .
\eeq
The dependence on $s$ can be easily isolated by a change of 
variables, so we can write:
\beq
\label{psmubis}
p(s ; \mu) \propto s 
\left( 1 + \frac{N^3 s^2}{2 \pi \mu} \right)^{-(1 + \mu/2)}  \; ,
\eeq
or alternatively 
\beq
\label{psmubisbis}
p(s ; \mu) \propto s \exp_{q_s}(-\alpha s^2) \; ,
\eeq
where 
\beq
q_s\equiv \frac{\mu+4}{\mu+2}\,\hspace{0.5cm}\mbox{and}\hspace{0.5cm}
\alpha\equiv\frac{N^3}{4\pi}\,\frac{\mu+2}{\mu}\;.
\eeq
The function of Eq.(\ref{psmubis}) (or Eq.(\ref{psmubisbis})) 
is {\em identical} in shape with the exact level-spacing distribution 
of the $2 \! \times \! 2$ qOE having the same $\mu$ (see appendix). 
Then, both distributions can be collapsed by a simple scaling 
of the arguments.
This curious result constitutes a generalization of Wigner's
surmise to qOE.

{\em Remark.} 
When analyzing spectral statistics it is usual to normalize 
energies so that the (local) average spacing is one (the spectrum
is ``unfolded"). This amounts to measuring energies in units of 
\beq
\Delta \equiv \int_0^\infty ds \, s \, p(s) \;.
\eeq
Note, however, that in qOE the first moment of $p(s)$ does not exist 
for $\mu \le 1$. 
In these cases, instead of $\Delta$, one may alternatively use the 
energy scale 
\beq
\widetilde{\Delta} \equiv 
\left[ \int_0^\infty ds \, s^{-1} \, p(s) \right]^{-1} \;.
\eeq
Due to level repulsion there is no singularity at $s=0$, and 
$\widetilde{\Delta}$ always exists, thus representing 
a characteristic energy of qOE. It is close to the inverse 
of the level density at $\varepsilon=0$:
\beq
\label{Delta}
\widetilde{\Delta} = \frac{2}{\pi \rho (0;\mu)} \; ,
\eeq
with $\rho(0;\mu)$ given in (\ref{rhozeromu}).

\section{Numerical results}
\label{sec4}

When thought of as clouds in $R^d$, via the map of Eq.~(\ref{map}),
both ensembles qOE and GOE are spherically symmetric. 
This means that qOE can be constructed just by rescaling the 
radii of all points in the GOE cloud \cite{lecaer99,delannay00}.
Thus, the construction of a qOE matrix ${\bf H}_1$
(with parameters $q$,$\sigma$,$N$) 
can be done in three steps.
(i) Construct a GOE matrix ${\bf H}_0$ of size $N \times N$. 
In this case matrix elements are independent and can be calculated 
using Eqs.~(\ref{marginal1}) and (\ref{marginal2}) with $q=1$. 
The radius of ${\bf H}_0$ is 
\beq
r_0 = \sqrt{\tr {\bf H}_0^2} \; .
\eeq
(ii) Choose a radius $r_1$ randomly according to the radial 
probability distribution 
${\cal P}(r_1,q,\sigma,N)$ of Eq.~(\ref{pofr}).
(iii) Define ${\bf H}_1$ as
\beq
{\bf H}_1 = {\bf H}_0 \frac{r_1}{r_0} \; .
\eeq
This is the recipe we followed for constructing qOE matrices.
(If, instead of being a random variable, $r_1$ is fixed,
we obtain a matrix belonging to FTE.) 
The only difficulty is to devise the random number generator,
especially when ${\cal P}(r)$ has very long tails. For this
purpose we used a combination of the {\em rejection method} 
and the {\em transformation method}, as explained in 
Ref.~\cite{recipes}.

\begin{figure}[!hbt]
\includegraphics[angle=-90,width=8.0cm]{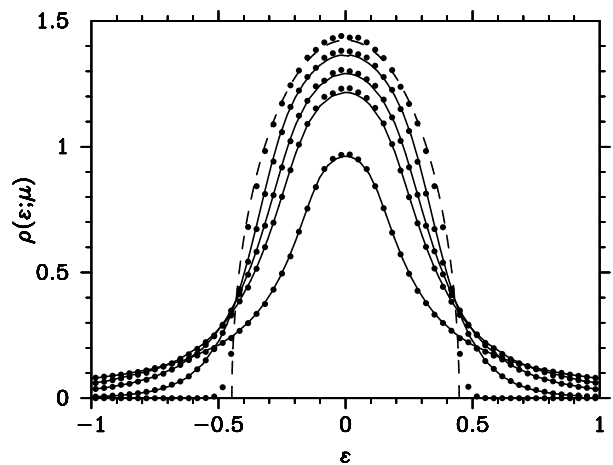} 
\caption{%
Density of states in the ensembles qOE (normalized to one). 
We compare histograms generated numerically (dots) with the 
theoretical result of Eq.~(\protect\ref{rhoemu}) (curves). 
Each histogram was generated from a set of $10^5$ matrices.
We used the following values of $\mu$: 
0.5, 1.5, 2.5, 6.0, $\infty$. 
Densities with larger $\mu$'s have larger values at $\varepsilon=0$ 
and decay faster.
In all cases $N = 20$ and $\sigma=1$.
The dashed line corresponds to the GOE semicircle ($N \to \infty$).}
\label{fig2}
\end{figure}

\begin{figure}[!hbt]
\includegraphics[angle=-90,width=8cm]{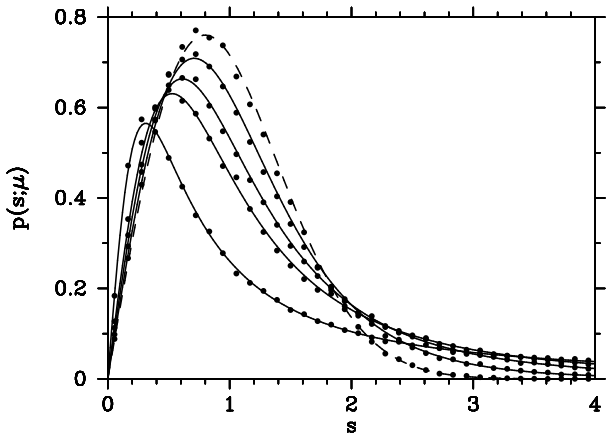} 
\caption{Level-spacing distribution in the ensembles qOE. 
We compare histograms generated numerically (dots) with the 
theoretical result of Eq.~(\protect\ref{psmubis}) (curves). 
Each histogram was generated from a set of $10^5$ matrices,
each matrix contributing one pair of levels.
We used the same values of $\mu$ as in Fig.~\protect\ref{fig1}.
Distributions with larger $\mu$'s have higher maxima and decay faster.
In all cases $N=20$ and $\sigma=1$.}
\label{fig3}
\end{figure}

In Fig.~\ref{fig2} we show histograms representing densities 
of states obtained from diagonalization of qOE matrices.
It is clear that they are very well described by the formula 
(\ref{rhoemu}), which was evaluated by direct numerical 
integration.
\vspace{1pc}
The statistics of level spacings is exhibited in Fig.~\ref{fig3}.
Histograms were obtained by binning data from $10^5$ matrices.
Each matrix contributed with the ``central" spacing between  
levels $\varepsilon_{N/2}$ and $\varepsilon_{N/2+1}$. 
The analytical curves are the $q$-distributions of Eq.~(\ref{psmubis}).
Again the agreement between theory and simulations is satisfactory.
In both figures we observe some small deviations, which may be attributed 
to the relatively small size of the matrices considered ($N$=20).
The values of $\mu$ were chosen in accordance to the following criterion.  
When $\mu=0.5$ all integer moments diverge. 
For $\mu=1.5$ ($\mu=2.5$) the first (second) moment exists but higher ones 
diverge. 
The case $\mu=6.0$ is intended to represent an ensemble qOE approaching GOE. 
\vspace{1pc}
%

\section{Conclusions}
\label{sec5}

This paper explores the possibility of using a modified maximum entropy 
method to construct random matrix ensembles. Whereas the usual logarithmic
entropy leads to ensembles characterized by exponential laws, 
the power-law entropy of  Eq.~(\ref{tsallis}) naturally produces ensembles 
with long-tailed distributions, (e.g., the so-called $q$-Gaussians \cite{tsallis}). These ensembles 
($q$OE) recover the Cauchy-Lorentz and Student's ones
for special choices of the parameters. 

Of course, the same families of ensembles can also be obtained from say the standard
maximum entropy approach, but at the expense of introducing a complicated 
constraint. 
A similar situation arises in the maximum entropy approach to anomalous
diffusion, where one can choose between an unappealing constraint  \cite{montroll83}
and a nonextensive entropy \cite{anomalous1d,anomalous}.
By sweeping the parameter $q$ one switches from sub- to superdiffusive regimes, 
$q=1$ giving normal diffusion \cite{anomalous1d,anomalous}.

In the present case, the entropic index $q$ controls the confinement, allowing to
access different universality classes. The anomalies we found can be associated to an 
effect of soft- (or weak-) confinement. However, the confining potential in qOE is many-body, as 
opposed to the more common single-particle confinement in standard Random Matrix 
Theories.

Gaussian ensembles are {\em ergodic} \cite{pluhar00}.
In Sect.~\ref{sec3} we explicitly used the fact that individual large qOE matrices 
have GOE statistics. This implies that qOE is {\it not ergodic} (if $\mu$ is finite).
In random matrix theory, non-ergodicity is considered to be a drawback because, it is argued,  
predictions (ensemble-averages) are compared with data obtained from a single 
system. We do not object to this reasoning, but just mention that in some cases 
empirical data are indeed extracted from ensembles of Hamiltonians 
\cite{haq82,levy93}.

\begin{acknowledgements}
We are grateful to C. Anteneodo, S. Ghosh, C. H. Lewenkopf, and 
A. M. Ozorio de Almeida  
for fruitful comments.
Partial financial aid from FAPERJ, 
CNPq, and PRONEX is gratefully acknowledged. 
\end{acknowledgements}

\appendix

\section{Generalized Wigner surmise}

Here we apply (\ref{relation}) to the level spacing 
distribution in qOE with $N=2$. We want to calculate 
\beq
\label{psqOE2}
p(s ; q, \sigma) \equiv 
\langle \delta \left[
\left(\varepsilon_2 - \varepsilon_1\right)({\bf H}) - s 
\right] \rangle_{\rm qOE} \; ,
\eeq
where $\varepsilon_1 < \varepsilon_2$ are the eigenvalues of $\bf H$.
First we need $p(s;r)$ for FTE. This is a known result 
\cite{delannay00}: 
\beq
\label{psFTE2}
p(s ; r) = \frac{s}{\sqrt{2}\;r} \; \frac{1}{\sqrt{2r^2-s^2}} \; 
\eeq
(if the argument of the root is positive; zero otherwise).
According to (\ref{relation}), $p(s;q,\sigma)$ is obtained by averaging
(\ref{psFTE2}) with the radial weight of qOE, Eq.~(\ref{pofr}) with
$d=3$. We show the result for the case $q>1$: 
\beq
\label{psq2}
p(s ; q, \sigma) 
\propto s
\exp_{q''} \left( -\lambda'' s^2 \right)  \; ,
\eeq
where
\beq
q'' = \frac{1+q}{3-q} \; , 
\eeq
and
\beq
\lambda'' = \frac{3-q}{4 \sigma^2(5/3-q)} \; .
\eeq
Equation (\ref{psq2}) plays the role of Wigner's surmise 
for qOE. It is more useful to rewrite (\ref{psq2}) in terms
of the parameter $\mu$ of Eq.~(\ref{mu}). The result
is
\beq
\label{psq2mu}
p(s ; q, \sigma) \propto s 
\left( 1 + \frac{3 s^2}{2 \mu} \right)^{-(1 + \mu/2)}  \; .
\eeq
In Sect.~\ref{sec3} this expression is compared with the spacing 
distribution for the large $N$ case (Eq. (\ref{psmubis})).



\begin{thebibliography}{99}
\bibitem{omission}
If desired, the normalization constants we omit here and there
can be easily calculated using any symbolic computing software.
\end{thebibliography}
\end{document}